\input cp-aa

\def\|{\partial}
\def\o {\over}
\def\s {\sigma}
\def\t {\tilde}
\def\r {\rho}
\def\f {\varphi}
\def\a {\alpha}
\def\Oo {\displaystyle}
\def\varkappa {{\scriptstyle\partial}\! e}
\def\d {\diff}
\def\i {\imag}

\MAINTITLE{Instability of high-frequency acoustic waves in accretion disks 
with turbulent viscosity}
\AUTHOR{A.V. Khoperskov, S.S. Khrapov}
\INSTITUTE{Department of Theoretical Physics, Volgograde State University, 
Volgograd 40068, Russia}
\DATE{Received  1997; accepted  1998}
\ABSTRACT{
Dynamics of linear perturbations in a differentially rotating accretion disk 
with non-homogeneous vertical structure is investigated.
It has been found that turbulent viscosity results in
instability of both pinching oscillations, and bending modes. 
Not only the low-frequency fundamental modes, but also the high-frequency
reflective harmonics appear to be unstable. 
The question of the limits of applicability of the thin disk model (MTD) 
is also investigated.
The insignificant distinctions in the dispersion properties of MTD 
and three-dimensional model appear for wave numbers $k \la (1\div 3)/h$ 
($h$ is the half-thickness of a disk). 
In the long-wavelenght limit, the relative difference between 
eigenfrequencies of the unstable acoustic mode in the 3D-model 
and the MTD is smaller than 5 \%.
}
\THESAURUS{02         
              (02.01.2;  
               02.08.1;  
               02.09.1;  
               02.20.1;  
	      }%
\maketitle
\MAINTITLERUNNINGHEAD{Instability of high-frequency acoustic waves in accretion 
disks }
%
\titlea{Introduction}

Accretion disks (AD) are important in many observable astrophysical objects 
(close binary systems, quasars, active galactic nuclei, young stars, 
protoplanet disks). 
It is usually assumed that the disk is geometrically thin, and that the disk 
has turbulent viscosity (Shakura \& Sunyaev 1973; Lightman \& Eardley 1974).  
The physical mechanism of turbulent viscosity can be connected with various 
instabilities. 

The basis for various viscous models of AD is the assumption, that the dynamic 
viscosity $\eta$ is caused by turbulence of medium and that 
$\eta \sim \r u_t \ell_t$ ($u_t$ is the characteristic amplitude of the most
large-scale turbulent velocity, $\ell_t$ is the spatial scale, $\r$ 
is the density) (Shakura \& Sunyaev 1973). 
In this connection the vital importance is given to the research of multimode
instabilities, which result in a complex perturbations structure. 
Different spatial and temporal scales may lead to the development of 
turbulence in the disk.

The thin disk model is the widely used for studies of the accretion disk 
dynamics.
The MTD involves averaging over $z$-coordinate of the three-dimensional 
hydrodynamic equations, if a number of additional conditions is fulfilled 
(Shakura \& Sunyaev 1973; Gor'kavyi \& Fridman 1994; Khoperskov \& Khrapov 1995). 
In the context of two-dimensional (in the plane of the disk) models without
magnetic field, four unstable modes of oscillations are present: 
two acoustic modes (Wallinder 1991; Wu \& Yang 1994; Khoperskov \& Khrapov 1995; 
Wu et al. 1995), a thermal mode and a viscous one (Lightman \& Eardley 1974; 
Shakura \& Sunyaev 1976; Szuszkiewicz 1990; Wallinder 1991; Wu \& Yang 1994). 
The instability growth rate of acoustic waves increases with the reduction of 
wavelength $\lambda$. 
However, the thin disk model imposes the restriction on wavelength, $\lambda 
\gg h$, and therefore it is necessary to consider AD $z$-structure for a
correct treatment when $\lambda \la h$. 
The thin disk model applies only to pinching oscillations. 
Then the perturbed pressure is a symmetric function, and the displacement
of gas does not move the mass centre in a disk with respect to a symmetry 
plane ($z=0$). 
Thus, bending oscillations (AS-mode) are excluded from consideration, for
AS-mode the perturbed pressure is antisymmetric function. 
Also, in two-dimensional models the high-frequency
(reflective)  harmonics with characteristic spatial scales in $z$-direction
$\la (0.5 \div 1) h$ cannot be investigated. 

In this paper we investigate the dynamics of acoustic perturbations taking 
into account the non-homogeneous $z$-structure of a viscous disk. 
Apart from the special problem of the limits of MTD applicability, 
main question is the existence of instabilities in short-wave region and, 
in addition, the stability of high-frequency harmonics.
In Sect. 2 we define the AD model and we choose the viscosity law. 
In Sect. 3 we consider the dynamics of linear acoustic perturbations, 
and formulate a mathematical problem of eigenfrequency determination 
for various unstable modes in the disk. 
Lastly in Sects. 4 and 5 we discuss results of the numerical solution 
of the boundary problem and summarize the main conclusions. 

\titlea{Model and basic equations}

We shall consider an axisymmetric differentially rotating gas disk 
in the gravitational field of a mass $M$. 
Without including self-gravity and relativistic effects, and adopting 
cylindrical coordinates we have:  
$$
\Psi(r, z) = - {GM\o (r^2 + z^2)^{1/2}}
\simeq - {GM \o r} + {1\o 2} \Omega_k^2 z^2
\,\,, \eqno(1)
$$
$G$ is the gravitational constant, $\Omega_k = \sqrt {GM/r^3} $ is Keplerian
angular velocity.

We shall use the axisymmetric hydrodynamic equations in view of viscosity.
The equations of motion and continuity have the form
$$
{ \d u \o \d t} - {v^2 \o r} = {1\o \r} \left(- {\| P\o \| r} + {\| r
\s_{rr} \o r \| r} - {\s_{\f \f} \o r} +
{ \| \s_{rz} \o \| z} \right) - {\| \Psi\o \| r} \,\,, \eqno (2) 
$$
$$
{ \d v \o \d t} + {u v \o r} = {1\o \r} \left({\| \s_{\f z} \o \| z} +
{\| r^2\s_{r \f} \o r^2 \| r} \right) \,\,, \eqno (3) 
$$
$$
{ \d w \o \d t} = {1\o \r} \left(- {\| P\o \| z} + {\| r \s_{rz} \o r \| r}
+ {\| \s_{zz} \o \| z} \right) - {\| \Psi\o \| z} \,\,, \eqno (4) 
$$
$$
{ \| \r \o \| t} + {\| (r \r u) \o r \| r} + {\| (\r w) \o \| z} = 0\,\,, 
\eqno (5) 
$$
where $\Oo {\d\o \d t} \equiv {\| \o \| t} + u {\| \o \| r} + w {\| \o \| z} $, $
{\bf \vec V} = (u, v, w) $ is the velocity, $P$ is the pressure, 
$\r$ is the volume density of matter in a disk, $\s_{ij} $ is the components 
of symmetric viscous stress tensor ($\s_{ij} = \s_{ji} $).

We shall add the thermal equation to the system of Eqs. (2)--(5) as
$$
{ \d S\o \d t} = {Q\o T} \,\,, \eqno (6)
$$
where $S$ is the entropy, $T$ is the temperature, and the variable $Q$ defines
sources of  heat. 

\titleb{Equilibrium model and viscosity law}

We shall assume that the equilibrium velocity in the disk has only $r$ and $\f$
components: $ {\bf \vec V}_0 = (U_0,\, V_0, \, 0)$.  
The expression for components of the viscous stress tensor may be written in
the following form:
$$
\s_{ij} = - \alpha_{ij} \, P \,\,, \qquad \a_{ij} = {\rm const} > 0 \,\,, 
\eqno (7) 
$$
where $i, j = (r,\f, z) $.
As without the account of the second (volume) viscosity the trace of the
viscous tensor is equal to zero ($\delta_{ij} \s_{ji} = 0$), thus 
$\a_{ii} = 0$ ($\s_{rr} = \s_{\f \f} = \s_{zz} = 0$).

The parameters $\a_{rz} $, $\a_{\f z} $ and $\a_{r \f} $ determine a level
of turbulence in the disk. 
Moreover $\a_{rz} $ and $\a_{\f z} $ are caused by a shear character of flow in
$z$-direction, and the value $\a_{r\f}$ is connected to differentiality of
rotation in the disk  plane and coincides with $\alpha$-parameter of the
standard theory of the disk accretion (Shakura \& Sunyaev 1973). 
For $\a_{ij}$ it is possible to write down:
$$
\a_{r\f} = - {\eta_0 \o P_0} \, l_\Omega \Omega \,,
\a_{rz} = - {\eta_0 \o P_0} \, {\| U_0 \o \| z} \,, 
\a_{\f z} = - {\eta_0 \o P_0} \, {\| V_0 \o \| z} \,, 
\eqno (8)
$$
here $l_\Omega = \| (\ln \Omega) / \| (\ln r) $.
One must be limited by the case of weak dependence of equilibrium
velocities ($U_0$ and $V_0$) on $z$-coordinate for correct transition to the
thin disk model. 
This is possible if the following conditions: $\a_{rz} \ll \a$ and $\a_{\f z}
\ll \a$ are fulfilled. 
Therefore it is possible not to take into account in the equations the terms,
containing $\s_{rz} $ and $\s_{\f z} $.
The conditions $U_0 \simeq \a (h/r)^2 V_0$ and $h/r\ll 1$ are carried out for
the majority of the models of accretion disks.
In view of the made above assumptions the equilibrium balance of the forces is
defined by the equations: 
$$ 
{ V_0^2 \o r} = {1\o \r_0} {\|
P_0 \o \| r} + {\| \Psi\o \| r} \,\,, \eqno (9) 
$$ 
$$
U_0 {\| r V_0 \o r \| r} = - {\a \o \r_0} {\| r^2 P_0 \o r^2 \| r} \,\,
, \eqno (10) 
$$
$$
{ \| P_0 \o \| z} = - \r_0 \, {\| \Psi \o \| z} \,\,. \eqno (11) 
$$ 

The equilibrium functions can be submitted in the self-similar form: 
$f_0 (r,z) = f_{01} (r) f_{02} (z)$, if it is assumed $h/r \ll 1$. 
The disk $z$-structure depends on the state equation and the energy  flux.
For the sake of simplicity we shall limit ourselves to a polytropic model
$$
{ \| \o \| z} \left\{P_0 (r, z) \o [\r_0 (r, z)]^n \right\} = 0\,\,. 
\eqno(12)
$$ 
Then for the pressure and density it is possible to write the solutions:  
$$ P_0 (r, z) = P_0 (r, 0) \, F (z)^a \,, \,
\r_0 (r, z) = \r_0 (r, 0) \, F (z)^b \,, \eqno (13) 
$$ 
where $F (z) = 1-z^2/h^2$, $a=n/ (n-1) $, $b=1/ (n-1) $, $n$ is polytropic
index, and  $h$ defines the disk boundary --- in the points $z=\pm h$ the
equilibrium pressure and density are equal to zero.  
We assume below that $h (r) = const$. 

The relation
$$ 
C_s^2 (r, z) = {\gamma P_0\o \r_0} =
C_s^2 (r, 0) \, F (z) \,, \,\, C_s^2 (r, 0) = {\gamma \o a} \,\Omega_k^2 h^2
\,, \eqno (14) 
$$
defines the adiabatic sound speed ($\gamma$ is adiabatic index). 
For the equilibrium velocities ($V_0$ and $U_0$) from Eqs. (9), (10)
taking into account Eqs. (13) and (14) we obtain:
$$
V_0 (r, z) = r\Omega (r, z) = \sqrt {r^2 \Omega_k^2 + l_P C_s^2/\gamma}
\,, \eqno (15) 
$$
$$
U_0 (r, z) = - {\a (2 + l_P) \o \gamma (1 + l_V)} \, {C_s^2 \o V_0}
\,, \eqno (16) 
$$
where $l_P = \| (\ln P_0) / \| (\ln r) $, $l_\rho = \| (\ln \rho_0) / \| (\ln
r) $, $l_V = \| (\ln V_0) / \| (\ln r) $.

\titlea{Linear analysis}

In the framework of standard linear analysis the pressure, density and 
velocity are represented as:  
$$
u = U_0(r, z) + \t u(r, z, t)\,, 
$$
$$
v = V_0(r, z) + \t v(r, z, t)\,,
$$
$$
w = \t w(r, z, t)\,, 
$$
$$
P = P_0 (r, z) + \t P (r, z, t) \,, 
$$
$$
\r = \r_0 (r, z) + \t \r (r, z, t)\,.
$$
In the linear approximation ($| \t f | \ll | f_0 | $) from Eqs. (1)--(5) the
linearized equations become: 
$$
{ \| \t u\o \| t} + U_0 {\| \t u\o \| r} + \t u {\| U_0 \o \| r} + \t w {\|
U_0 \o \| z} - {2V_0 \o r} \, \t v =  
$$
$$
- {1\o \r_0} \left ({\| \t P \o \| r} - {\t \r \o \r_0} \, {\| P_0 \o \| r} 
\right) \,, \eqno (17)
$$
$$
{ \| \t v \o \| t} + U_0 {\| (r \t v) \o r \| r} + \t u {\| (r V_0) \o r \| r}
+ \t w {\| V_0 \o \| z} = 
$$
$$
- {\a \o \r_0} \left( {\| (r^2 \t P) \o r^2 \| r} - 
{ \t \r \o \r_0} {\| (r^2 P_0) \o r^2 \| r} \right) \,, \eqno (18)
$$
$$
{ \| \t w \o \| t} + U_0 {\| \t w \o \| r} 
= - {\| \t P \o \r_0 \| z} - g {\t \r \o \r_0} \,, \eqno (19)
$$
$$
{ \| \t \r \o \| t} + {\| \o r \| r} \left [r (U_0 \t \r + \r_0 \t u) \right]
+ {\| (\r_0 \t w) \o \| z} = 0 \,, \eqno (20)
$$
where $g \equiv \| \Psi /\| z$.

As we shall study the dynamics of acoustic oscillations, it is possible 
to set $Q = 0$. Then the thermal Eq. (6) becomes:
$$
{ \| \t S \o \| t} + U_0 {\| \t S \o \| r} + \t u {\| S_0 \o \| r} 
+ \t w {\| S_0 \o \| z} = 0 \,, \eqno (21)
$$
here $S_0 = c_V \ln (P_0 / \rho_0^\gamma) $ is the entropy of equilibrium gas.
We shall exclude $\t S$ from Eq. (21) with the help of the equation of state 
$\t S = \t S (\t P, \t \rho) $. 
In the linear approximation we obtain:
$$
\t S = \left ({\| S\o \| P} \right)_\rho \, \t P + 
\left ({\| S\o \| \rho} \right)_P \, \t \rho = 
c_V {\t P\o P_0} - c_P {\t \rho \o \rho_0} 
\,, \eqno (22)
$$
where $c_V = T (\| S/ \| T)_\rho$, $c_P = T (\| S / \| T)_P$ are specific heats at
constant density and pressure.

Short-wave approximation in a radial direction ($kr \gg 1$, $k$ is radial 
wave number) makes it possible to write the solution as: 
$$
\t f (r, z, t) = \hat f (z) \, \exp\{\i k r - \i \omega t \} \,\,,\eqno (23)
$$
here $\omega$ is complex eigenfrequency. 

Taking into account Eqs. (22) and (23), the system Eqs. (17)--(21) is reduced 
to two ordinary differential equations for amplitudes of the perturbed pressure
$\hat P (z) $ and the material displacement $\hat \xi (z) $ from an equilibrium
position:
$$
{ \d \hat \xi \o \d z} = {D_1 \o \hat \omega q^2 C_s^2}
{ \hat P\o \r_0} +
\left\{{k D_2 \o \hat \omega q^2} + {g\o C_s^2} \right\}
\,\hat\xi 
\,, \eqno (24)
$$
$$
{\d \hat P \o \d z} = \r_0 \left\{\hat\omega^2- {g\o \gamma} D_3 \right\} 
\hat\xi - {g\o C_s^2} \left\{1 + D_4 \right\} \hat P
\,, \eqno (25)
$$
where $\hat \omega = \omega - kU_0$,  
$q^2 = \hat \omega^2 - \varkappa^2$,  
$\varkappa = \Omega \sqrt {2 (2 + l_\Omega)} $ is the epicyclic frequency, 
$\Oo D_1 = - \hat\omega^3 + \hat\omega (\varkappa^2 + k^2 C_s^2) + 
2\i\a\Omega k^2 C_s^2 $, 
$\Oo D_2 = \i\omega F_V - 2\Omega F_U$,
$\Oo D_3 = S_z - {\i S_r \o \hat \omega q^2} \, 
(\hat \omega F_U + 2\i\Omega F_V)$,
$\Oo D_4 = {\i k S_r \o \gamma \hat \omega q^2} \, (\hat \omega + 2\i\a \Omega)$,
$\Oo F_V = {l_P C_s^2 \o r \gamma^2} \, S_z - 2\Omega V_0'$,
$\Oo F_U = {\a (2 + l_P) C_s^2 \o r \gamma^2} \, S_z + 
{ \varkappa^2 \o 2\Omega} U_0'$, 
$\Oo S_z = {1\o c_V} {\| S_0 \o \| z} = (\ln P_0) ' - \gamma (\ln \rho_0) ' $,
$\Oo S_r = {1\o c_V} {\| S_0 \o \| r} = (l_P - \gamma l_\rho) /r$, 
the stroke means differentiation with respect to $z$-coordinate 
($f'\equiv \| f / \| z$), 
$\hat \xi$ is the complex amplitude of material $z$-displacement from
equilibrium state, and the following is true
$
\t w = \d \t \xi / \d t = -\i\hat \omega \,\t \xi \,.
$

Boundary conditions must be added to Eqs. (24) and (25). 
In view of the symmetry, it is natural to consider two types of
oscillation: 1) the symmetric oscillations ($\hat \xi (z) = -\hat \xi (-z) $
or $\hat P (z) = \hat P (-z) $), and therefore  
$$
\hat \xi (0) = 0 \qquad {\rm or} \qquad {\d \hat P \o \d z} \Big |_{z=0} = 0
\,, \eqno (26)
$$
(such oscillations correspond to the pinch-mode or S-mode); 
2) the antisymmetric oscillations ($\hat \xi (z) = \hat \xi (-z) $ or $\hat P
(z) = -\hat P (-z) $), and therefore  
$$
\hat P (0) = 0 \qquad {\rm or} \qquad {\d \hat \xi \o \d z} \Big |_{z=0} = 0
\,, \eqno (27)
$$
that correspond to the bending oscillations or AS-mode.
On the disk unperturbed surface the following condition should be
fulfilled:
$$
\hat P (h) + {\| P_0 \o \| z} \Big |_{z=h} \hat \xi (h) = 0 \,. \eqno (28)
$$

By solving the system of Eqs. (24), (25) with the boundary conditions 
Eqs. (28) and (26) (or (27)) (boundary problem), we find the eigenvalues 
of complex frequency $\omega$ for the given distribution of equilibrium
parameters along the vertical coordinate in the disk. 
A positive imaginary part of the frequency (growth rate) means that the
eigen-mode is unstable. 
The above-stated outline will have the name "3D-model ", and the
model of thin disk --- "2D-model ". 

\titleb{Dispersion relation}

For a homogeneous $z$-distribution of the equilibrium quantities, the system 
of Eqs. (24) and (25) is reduced to the following dispersion relation: 
$$
\omega^4 - \omega^2 [\varkappa^2 + C_s^2 (k^2 + k_z^2)] - 2\i\omega 
\a \Omega C_s^2 k^2 + \varkappa^2 C_s^2 k_z^2 = 0\,, \eqno (29)
$$
here $k_z$ is the wave number in $z$-direction.
It is necessary to stress that Eq. (29) at $k_z=0$ coincides with the
dispersion relation earlier obtained for 2D-model in case of replacement
$\gamma$ by the flat adiabatic index 
$\Oo \Gamma = \Gamma_1 = (3 \gamma -1)/ (\gamma + 1)$ 
(Wallinder 1991; Khoperskov \& Khrapov 1995). 
Kovalenko and Lukin (1998) have obtained the more accurate formula for the 
flat adiabatic index 
$
\Oo \Gamma = \Gamma_2 = (3 \gamma - 1 - \gamma \delta) / (\gamma + 1- \delta)
$ 
($\delta = \varkappa^2 / \Omega_k^2$).
This relation takes into account Keplerian rotation of a thin disk. 
The adiabatic sound speed in the disk plane should be determined
by the following expression: 
$$
C_s^2 = \Gamma \, {\int_0^h P (z) \, \d z \o \int_0^h \r (z) \, \d z} \,.
\eqno (30)
$$
Thus, taking into account Eq. (30), the dispersion relation Eq. (29) 
in the case of $k_z=0$  describes dynamics of perturbations within 
the limits of flat model (MTD). 
Obviously, these oscillations correspond to S-mode. 

In the general case when $\a > 0$, this dispersion relation gives 
two unstable acoustic branches of oscillations and two damping modes 
(viscous and thermal ones). 
The damping of the viscous or thermal modes depends on the fact that we don't
take into account dissipation and radiative processes in the thermal equation. 
If these factors are taken into account, then the viscous and thermal
low-frequency oscillatory branches can be unstable (Shakura \& Sunyaev 1976),
but this effect will not change the dispersion properties of sound waves.

\titlea{Results}

It is convenient to characterize the properties of the considered
acoustic oscillations in the disk by the dimensionless frequency 
$W = \omega /\Omega$ and the dimensionless wave number $kh$.
We shall define the basic model as follows: 
$\a = 0.2$, $\gamma = 5/3$, $h/r= 0.05$, $l_P=-3/2$, $l_\rho=-1/2$, $l_V =
-1/2$ ($l_\Omega = -3/2$), $n = 5/3$. 
If it is not specified otherwise, the parameters take these values.

\titleb{Fundamental S-mode in the 2D- and 3D-models}

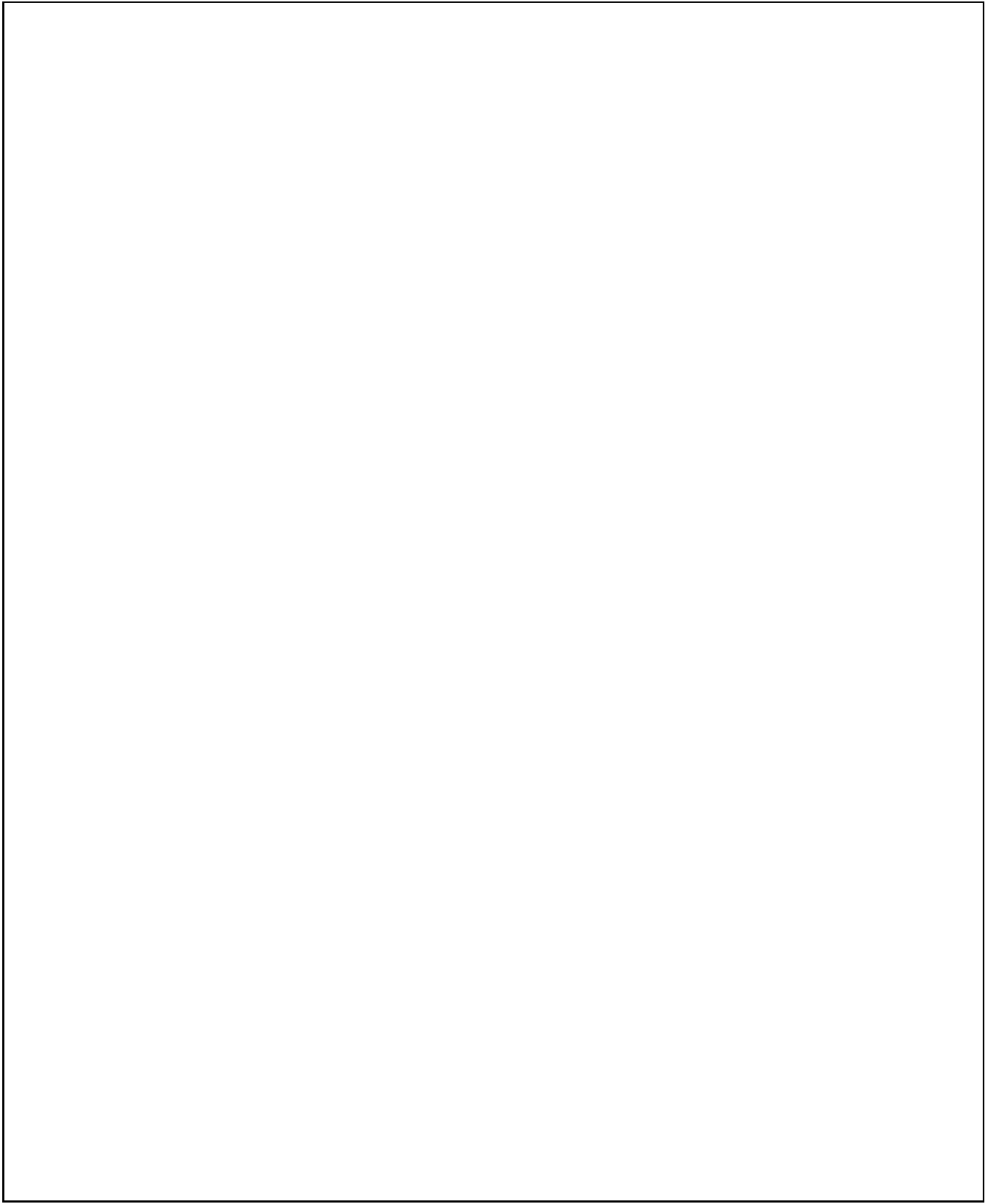
\begfigwid 22 cm {\figure{1a-f}%
             {The dependence of the dimensionless eigenfrequency $W$ on
              the  dimensionless wave number $kh$ for different values 
              of $n$ (1 --- $n=5/3$, 2 --- $n=1.2$) (see {\bf a-d}). 
              The dependence of $ \| W / h \| k $ on $kh$ 
              for S-mode (see {\bf e,f}). 
	      The solid lines correspond to 3D-model, 
	      dotted lines --- 2D-model at $\Gamma = \Gamma_1$,  
	      long-dashed lines --- 2D-model at $\Gamma = \Gamma_2$.  
              }}%
\endfig

Our analysis confirms the good agreement between the thin disk model and the
results of the 3D-problem solution in the case of the dissipational 
acoustic instability. 
The eigenfrequencies of oscillations obtained from Eq. (29) at $k_z=0$
by taking into account Eq. (30) and from the solution of Eqs. (24), (25)
practically coincide in the range $kh \la 3$ (see Fig. 1a,b). 
The visible differencies occur when $kh \ga 3$.
We emphasize that in this region the formal condition of applicability 
of the MTD ($kh \ll 1$) is obviously broken. 
In Fig. 1e,f, the dependencies of group velocity 
$v_g = {\rm Re}\left( \| \omega / \| k \right)$ and 
${\rm Im}\left( \| \omega / \| k \right)$ 
on the wave number $k$ for the 3D-model 
and the 2D-model at $\Gamma = \Gamma_1$ and $\Gamma = \Gamma_2$ are shown. 
In the case of small values of $n$, the divergence occur at large $kh$ as 
Fig. 1a,b shows. 
The reason is that the characteristic scale of inhomogeneity in the vertical 
direction increases with the reduction of $n$. 
The growth rate and the phase velocity of perturbations in the framework 
of the 3D-model is less than for the thin disk model (2D-model) in the 
short-wavelength region. 
This effect is caused by the inhomogeneous distribution of the equilibrium
quantities in $z$-direction and by the transverse gravitational force. 
When the parameter $\alpha$ increases the growth rates linearly grow, and the
value of the wave number $k$, at which the distinctions between the models
appear, does not depend on $\alpha$.
The considered low-frequency mode exists both in the 2D-model and in 
the 3D-model, and it weakly depends on $z$-structure. 
We therefore call it the {\it fundamental mode}. 
In the case $\alpha \ll 1$, the expression for frequency is easily 
obtained from Eq. (29) as:
$$
\omega = \pm \sqrt{ \varkappa^2 + k^2 C_s^2 }+ i\alpha {\Omega C_s^2 k^2 \o
\varkappa^2 + C_s^2 k^2}\,.
$$
This approximation is exact enough.
Our results shows, that the MTD adequately describes dynamics of perturbations 
with characteristic spatial scale $\lambda \sim 2h \div r$ in the disk plane. 

The differential rotation ($\Oo l_\Omega=- {d \ln \Omega \o d \ln r} > 0$) 
and a dependence of dynamic viscosity $\eta$ on thermodynamic parameters 
(for example, dependence on density and temperature) are responsible 
for the instability. 
A simple model can demonstrate this effect. 
The last term in Eq. (3) (i.e., containing $\sigma_{r\f}$) is a
source of the  instability of acoustic modes.
The expression $\eta = \sigma\nu$ ($\sigma$ is the surface density, 
$\nu$ is the kinematic viscosity) is usually used in the 2D-model and 
for the sake of simplicity we assume here that $\nu $ is constant. 
Then only a perturbation of surface density $\tilde\sigma$ generates 
perturbation of dynamic viscosity $\tilde \eta$.
Now let us consider how a viscous force, which is caused by $\tilde\eta$, 
leads to  an amplification of the amplitude of sound wave in a disk plane. 
The evolution equation for the surface density, without including 
Coriolis's force, is 
$ \Oo {\| \o \| t} \left [{\| ^2 \tilde \sigma \o \| t^2}
- C_s^2 {\| ^2 \tilde \sigma \o \| r^2} \right] = 
A {\| ^2 \tilde \sigma \o \| r^2} $,
where $A = 2\Omega^2 l_\Omega \nu_0 > 0$.
This equation is linear and the appropriate dispersion relation is 
$\omega [\omega^2 - C_s^2k^2] =\i Ak^2$, which means that acoustic waves are 
unstable with the growth rate $\Oo \Im (\omega) \simeq {\i \o 2} {A\o C_s^2}$ 
and entropy oscillations damping ($\Im (\omega) \simeq - \i A/C_s^2 < 0$).
These estimaties can only be used in the short-wavelength limit, because 
we have neglected epicyclic oscillations.

\titleb{Fundamental AS-mode}
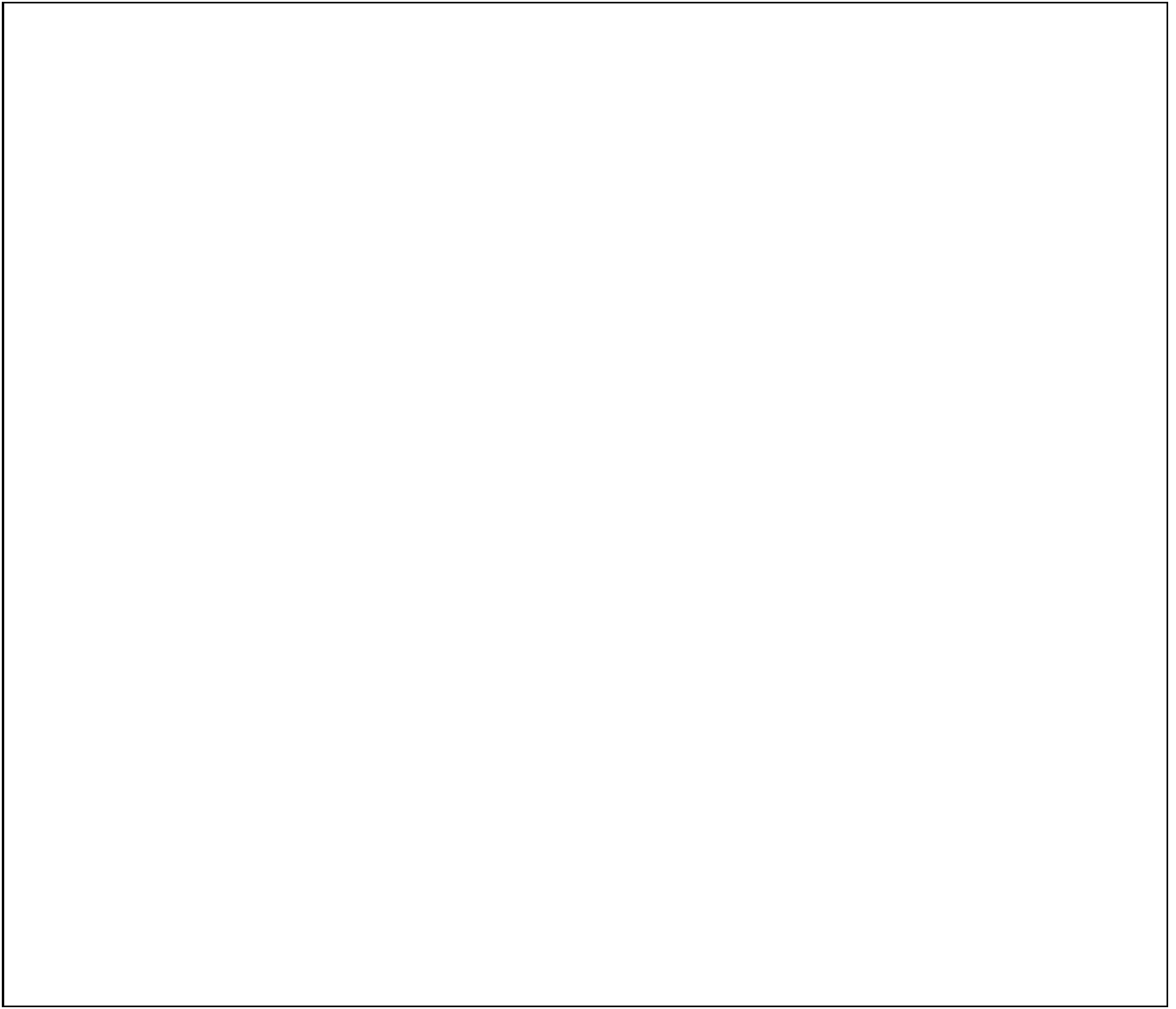
\begfigwid 15.5cm {\figure{2a-d}%
               {The dependence of $W$ on $kh$ for S- and AS-modes for
              various values of $\a$. The solid line corresponds to 
              $\a=0.2$, the dotted line --- $\a = 0.1$. The numbers 
              of harmonics $j$ are specified in Figure.  
	      }}%
\endfig

We shall consider low-frequency bending oscillations for which Eq. (27) 
is fulfilled.
In a longwavelength limit both symmetric and antisymmetric perturbations 
have identical frequency $\omega^2 \simeq \varkappa^2 = \Omega_k^2$ in the 
case of Keplerian disk. 
This result can easily be obtained from Eq. (29) when $k=0$.
In case of the adiabatic model two other branches of oscillations 
($\omega^2=(C_s k_z) ^2$) always damp at $k > 0$ and we do not examine them.
The dispersion behaviour of S- and AS-modes is very similar 
(Fig. 1a-d, solid lines).
For the fundamental bending mode it is possible to set $k_z h \simeq \pi /2$ 
in Eq. (29) and for this case the dispersion curves on Fig. 1c,d are 
shown as dotted lines.
The exact solution in the context of 3D-model shows visible difference from 
the result obtained from Eq. (29). 
In the range $kh \la 0.1$ these differences are caused by the following factors: 
1) the 3D-model includes radial inhomogeneities of equilibrium quantities; 
2) the Eq. (29) does not take into account vertical inhomogeneity of disk. 
The obtained result is very unexpected and significant, since we used 
Eq. (29) beyond the limit of 2D-model approach.

The physical reason for AS-mode instability is similar to the case of
S-oscillations.

\titleb{High-frequency S- and AS-modes}
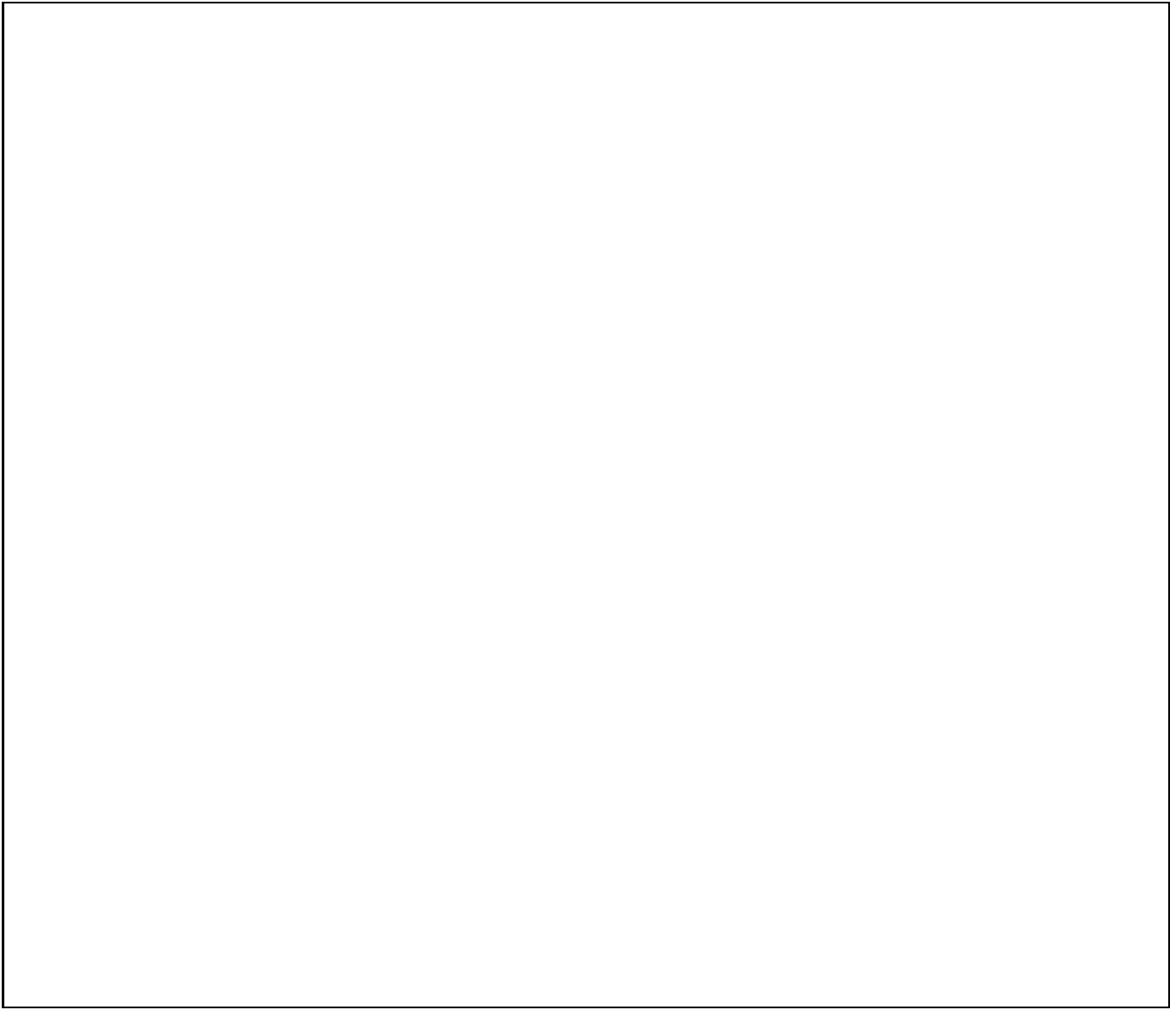
\begfigwid 15.5cm {\figure{3a-d}%
                { The dependence of $W$ on $kh$ for fundamental mode
               ($j=0$) and for high-frequency one with $j=3$ for 
               various values of $n$. Solid line --- $n= 5/3$, 
               dotted line --- $n=1.2$, dashed line --- $10$. 
               The numbers of harmonics $j$ are specified in Figure.  
	       }}%
\endfig

Besides the fundamental S-mode, any number of unstable harmonics can be 
generated in addition. 
These harmonics differ from each other by the number of nodes of the perturbed 
pressure across the disk plane. 
The following estimate of the effective wave number in the $z$-direction 
can be made:   
$$
k_z h \simeq \pi j \,\, ({\rm S-mode}) \qquad k_z h \simeq \pi (j + 1/2) \,\,
( {\rm AS-mode}) \,\,, 
$$
where $j$ is the number of the harmonic.
The fundamental ($j=0$) and reflective ($j > 0$) harmonics exist for both
symmetric (S-) and antisymmetric (AS-) modes.

The dependencies of eigen-frequency $\omega$ on the radial wave number $k$ for
fundamental mode $j=0$ and first four reflective harmonics $j=1,2,3,4$ are
displayed in Fig. 2a-d. 
Both the pinch-oscillations and the bending modes are unstable ($ \Im
(\omega) > 0$).  
The imaginary part of the frequency grows with $k$ and reaches a maximum 
at some value. 
The maximum of $\Im (\omega)$ moves to region of shorter wavelengths 
when the harmonic number $j$ grows, and the value increases with
reduction of the characteristic scale of perturbations in $z$-direction. 
As indicated by Fig. 2a-d, the growth rate increases with $\a$, while 
the perturbations phase velocity $\Re (\omega)/k$ does not depend on $\a$.
It should be noted that for very short-wave perturbations ($kh \ga 10$), 
the damping out of oscillations due to the presence of a perturbed velocity
gradient in the viscous stress tensor can be of decisive importance 
(Khoperskov \& Khrapov 1995). 
This factor can be essential for small-scale waves (as on $r$- and on 
$z$-coordinate) and it can result in complete stabilization.

The vertical structure of the disk in the 3D-model is determined 
primarily by parameters $\gamma$ and $n$.  
In the case when $n=\gamma$, entropy does not vary along the $z$-coordinate, 
i.e. the disk is obviously stable against convective $z$-motions.
When $n > \gamma$ we have $\| S_0/\| z<0$ and the conditions for convective
instability are fulfilled, in the opposite case $n < \gamma$ and 
$\| S_0 / \| z>0$, so convection can not appear. 
In Fig. 3a-d, the dispersion curves for various values of $n$ ($n > \gamma$, 
$n < \gamma$ and $n=\gamma$) for fundamental modes ($j=0$) and for the third
reflective harmonic ($j=3$) are shown.
The phase velocity of fundamental S- and AS-modes weakly depends on the
parameter $n$ (see Fig. 3b,d), as these perturbations are most longwave in
$z$-direction. 
The adiabatic sound speed in disk $C_s$ grows with $n$ (see Eq. (14)), 
and since for the high-frequency reflective harmonics 
$k_z \sim | d \hat \xi /\hat \xi dz | \ne 0$, the phase velocity of 
perturbations with $j > 0$ increases also (see Fig. 3b,d). 

The dispersion properties of the acoustic perturbations do not depend on the 
values of parameters $l_P$ and $l_\rho$ in the range $kh >0.2$.
The reason for this is that the radial gradients of equilibrium pressure and
density give a small contribution to the equilibrium balance in the case of 
a thin disk.

\titlea{Discussion and conclusions}

Acoustic waves in a differentially rotating gaseous disk can play an important 
role for understanding the nature of turbulent viscosity in accretion disks. 
The amplitude of small-scale (in $r$ and in $z$) waves grows most rapidly. 
Such instabilities do not destroy the initial flow at a nonlinear stage,  
but can effectively make the disk matter turbulent, and in turn, the 
arising turbulent viscosity generates unstable sound modes. 
As a result, a self-consistent regime with turbulent viscosity arises. 
Beside the dissipational mechanism analysed above, the development of global
(covering practically the whole disk radially) resonant Papaloizou-Pringle 
modes (Papaloizou \& Pringle 1987; Savonije \& Heemskerk 1990), and also 
resonant amplification of acoustic oscillations in the regime of double-flow 
accretion (Mustsevoj \& the Khoperskov 1991), and in the case of 
disk accretion onto a magnetized compact object (Hoperskov et al. 1993) 
may be important for understanding turbulent viscosity in the disk. 
It is remarkable that in all cases the characteristic time scale does not 
exceed the dynamical time scale ($\tau = 1/ \Im (\omega) \sim \Omega^{-1} $), 
and that all values are of the same order. 

We have demonstrated the possibility of unstable high-frequency acoustic 
waves in a differentially rotating gaseous disk.  
They exist in the system for a limited period, and the presence of a positive 
growth rate does not mean that the perturbations must reach a nonlinear stage. 
The perturbations leave the disk with a speed 
$\Oo{\| \omega /\|k} \sim C_s kh /\sqrt {1 + k^2h^2} < C_s $ 
and the characteristic life-time in the disk is $\sim r/h \sim 10^2$ disk 
rotation periods.
In view of the derived estimation $ \Im (\omega) \sim 0.1 \Omega$ for $kh > 1$ 
we have the essential growth of wave amplitude ($\exp \{0.1 r/h \} $).
Nonaxisymmetric perturbations may stay in the system for a longer time 
and reach a non-linear stage. 

The following are the main conclusions:

\item{1.} The linear wave dynamics of a thin disk model is compared to
an exact solution, which takes into account vertical  structure. 
Our analysis shows that not only the longwave perturbations, but also rather
short waves can be considered.
A small quantitative difference occur for a wavelength $\lambda < 2\pi h$, and
even in the case of $\lambda > 2\pi h$ the dispersion properties remain
qualitatively similar.
At $kh\le 1$ eigenfrequencies differ less than on 5\%.  

\item{2.} In addition to the fundamental dissipational unstable sound mode in
MTD, an arbitrary number of high-frequency unstable harmonics in the case of
pinch-oscillations is found.
These harmonics have a different vertical structure. 
The  waves with small scales along $z$-coordinate have the maximum growth rate
at short wavelengths in the $r$-direction.
The differential rotation and the variable dynamic viscosity cause instability
of all oscillation branches, i.e., a dissipational mechanism lead to the
perturbation growth.

\item{3.} Taking into account the vertical structure of disk we studied new
bending modes, which cannot be investigated in the context of MTD.
The bending oscillations, as well as the pinch-wave, are unstable. 
The physical mechanism of instability and the dispersion properties are
similar to the case of pinch-oscillations.
The significant feature of all considered unstable acoustic modes is the fact
that spatial scales of perturbations differ from each other, but characteristic
growth rates have the same order of magnitude.

\acknow{
     The authors thank I.G.~Kovalenko, V.M.~Lipunov, V.V.~Mustsevoj and 
     M.E.~Prokhorov for fruitful discussions and valuable comments.  
     This research is partially supported by Fund INTAS, grant number 95-0988.}
\begref{References}

   \ref Gor'kavyi N.N., Fridman A.M. Physics of planetary rings: Celestial mechanics of continuous medium. --- M.: Nauka, 1994. --- 348 p. 

   \ref Hoperskov A.V., Mustsevaya Yu.V., Mustsevoj V.V., Astronomical and astrophysical transactions, 1993, 4, 65

   \ref Khoperskov A.V., Khrapov S.S., Astronomy Letters, 1995, 21, No.~3, 347

   \ref Kovalenko I.G., Lukin D.V., Astronomy Reports, 1998, in press

   \ref Lightman A.P., Eardley D.M., ApJ, 1974, 187, L1 

   \ref Mustsevoj V.V., Khoperskov A.V., The letters in astronomical magazine, 1991, 17, 281

   \ref Papaloizou J.C.B., Pringle J.E., MNRAS, 1985, 213, 799

   \ref Papaloizou J.C.B., Pringle J.E., MNRAS, 1987, 225, 267

   \ref Savonije G.J., Heemskerk M.N.M., A\&A, 1990, 240, 191

   \ref Shakura N.I., Sunyaev R.A., A\&A, 1973, 24, 337 

   \ref Shakura N.I., Sunyaev R.A., MNRAS, 1976, 175, 613

   \ref Szuszkiewicz E., MNRAS, 1990, 244, 377

   \ref Wallinder F.H., MNRAS, 1991, 253, 184

   \ref Wu X-B., Yang L-T., ApJ, 1994, 432, 672

   \ref Wu X-B., Li Q-B., Zhao Y-H., Yang L-T., ApJ, 1995, 442, 736

   \ref Wu X-B., Li Q-B., ApJ, 1996, 442, 736

\endref
\bye